\documentclass[twocolumn,aps,prl,superscriptaddress]{revtex4}

\usepackage{amsmath}
\usepackage{graphicx}
\usepackage{MnSymbol}

\usepackage[normalem]{ulem}
\usepackage{color}

\begin{document}

\title{Electroadhesion for soft adhesive pads and robotics: theory and numerical results}

\author{Bo N.J. Persson}
\affiliation{Peter Gr\"unberg Institut-1, FZ-J\"ulich, 52425 J\"ulich, Germany (email: b.persson@fz-juelich.de)}
\address{MultiscaleConsulting, Wolfshovener Str 2, 52428 J\"ulich, Germany}
\author{Jianglong Guo}
\affiliation{SoftLab, Bristol Robotics Laboratory, University of Bristol, Bristol, UK (email: J.Guo@bristol.ac.uk)}

\begin{abstract}
Soft adhesive pads are needed for many robotics applications, and one approach is based on electroadhesion. Here we present a general analytic model and numerical results for electroadhesion for soft solids with arbitrary time-dependent applied
voltage, and arbitrary dielectric response of the solids, and including surface roughness. 
We consider the simplest coplanar-plate-capacitor model with a periodic array of conducting strips located close to the surface of the adhesive pad, and discuss the optimum geometrical arrangement to obtain the maximal electroadhesion force. For surfaces with roughness the (non-contact) gap between the solids will strongly influence the electroadhesion, and we show how the electroadhesion force can be calculated using a contact mechanics theory for elastic solids. The theory and models we present can be used to optimize the design of adhesive pads for robotics application.
\end{abstract}

\maketitle

\pagestyle{empty}


{\bf 1 Introduction}

The Danish engineers Alfred Johnsen and Knud Rahbek discovered a century ago that an attractive force
occurs between two contacting materials when there is an electrical potential difference between them \cite{[1]}. The
term ``electroadhesion'' was coined to denote this electrostatic attraction \cite{[1],[2]}.
The electrical attraction between a charged surface and a human finger was discovered by Johnsen and Rahbek.
In 1953 Mallinckrodt et al.\cite{[3]} applied an alternating voltage to insulated metal electrodes
and observed an alternating electrostatic force that periodically attract and release the finger from the surface;
this is now denoted electrovibration \cite{[4],[5],[6],[7]}, and forms the basis for electroadhesion based haptic devices such as touchscreens and tactile displays. For these applications, tactile sensations are produced by the application of a voltage to the
conductive layer of an insulated haptic device such as a touchscreen, inducing
electroadhesive forces between the device and the approached user finger. If the applied electric voltage is modulated
in time the friction force acting on the finger will generate sensorial experiences \cite{[4],[5],[6],[7]}.

\begin{figure}
\includegraphics[width=1.0\columnwidth]{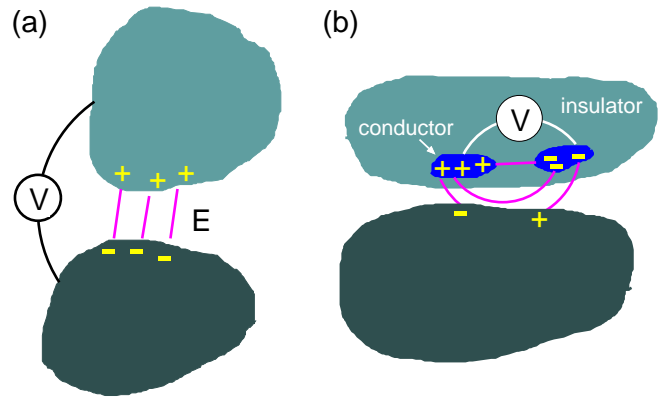}
\caption{\label{TwoWays.eps}
An attractive electrostatic force can occur between two solid objects if (a) an electric potential difference
occurs between the bodies, or if (b) an electric field occurs outside of one of the bodies due to an electric potential
difference between two conductive regions inside the body.}
\end{figure}

The Johnsen-Rahbek effect is due to the electrostatic attraction between the polarization charges on two solids resulting from an applied electric potential, see Fig. \ref{TwoWays.eps}(a). A related application is electrostatic chucks \cite{[8],[9]}, which have been utilized for various material handling tasks, such as wafer pick-up and place tasks \cite{[8]}. In this case an electric potential difference occurs between two metallic electrodes attached to the same object as in Fig. \ref{TwoWays.eps}(b). Electrostatic chucks are typically made from elastically stiff materials with very flat and smooth surfaces, which are useful for moving object with flat and very smooth surfaces like silicone wafers. However, in robotic applications the solid objects to be manipulated often have complex shapes \cite{[10],[11]} and large surface roughness. In these cases the adhesive pad must be built from an elastically soft material in order to increase the contact area and hence the friction force.  

Modeling of electroadhesive forces has been based on the parallel-plate-capacitor structure \cite{[12],[13],[14],[15]} when an electroadhesion pad is contacting a conductive or semi-conductive substrate
material, and the coplanar-plate-capacitor structure \cite{[16],[17],[18]} when an electroadhesion pad is contacting an insulating
substrate material.

\begin{figure}
\includegraphics[width=0.85\columnwidth]{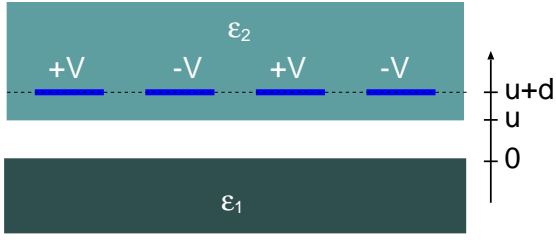}
\caption{\label{pic1}
Cross-sectional view of an ectroadhesion set-up with conducting strips located in the surface 
region of the adhesive pad, a distance $d$ above the surface.
By applying different electric potential to different conducting strips, an electric field is
generated and extends outside the adhesive pad and can polarize the substrate and induce an
attractive force between the two solids [see Fig. \ref{TwoWays.eps}(a)]. 
The substrate has the dielectric function $\epsilon_1$ and the electric
resistivity $\rho$ and the adhesion pad dielectric is a perfect insulator with the dielectric function $\epsilon_2$.
We first assume that there is a uniform air gap of width $u$ between the solids.
}
\end{figure}

In this paper we consider the simplest electroadhesion structure shown in Fig. \ref{pic1},
where the electroadhesion pad is made of a periodic array of coplanar electrodes
embedded in a soft dielectric.
By applying different electrical potentials to the two electrodes, an electric field is generated
which extends outside of the pad, polarizing the dielectric substrate and thus inducing an electroadhesive force
between the two solids \cite{[19],[20]}. This normal force moves the soft electroadhesion pad closer to the
substrate by squeezing the air gap and elastomeric coating of the pad. This will increase the sliding
friction force, which is important when using the electroadhesion pad in the tangential direction.

In this study we assume that the temperature and humidity are constant, and ignore the 
influence of contaminates, the Van der Waals force, and other attractive force fields which may act 
between the two solids.

\vskip 0.3cm
{\bf 2 Theory}

\begin{figure}
\includegraphics[width=0.85\columnwidth]{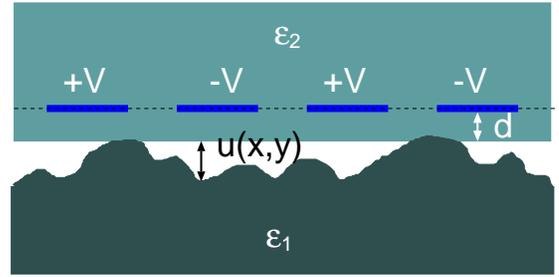}
\caption{\label{pic2}
Electroadhesion set-up with conducting strips located in the surface region of the adhesive pad.
The substrate has surface roughness and the contact between the solids is incomplete.
By applying different electric potential to different conducting strips, an electric field is
generated which extends outside the adhesive pad and can polarize the substrate and induce an
attractive force between the two solids.
}
\end{figure}

We first calculate the electroadhesion force when the surface separation $u$ 
is constant (see Fig. \ref{pic1}). Next we include surface roughness (see Fig. \ref{pic2}) 
and present a mean-field theory of electroadhesion. 

\vskip 0.2cm
{\bf 2.1 Electroadhesion with a constant air-gap}

The electroadhesive force can be derived by the Maxwell stress tensor 
method\cite{add,[13],[14],[15],[16],[17],[18]}. 
We calculate the electroadhesive force between the two solids when the surfaces are separated by a distance $u$ (see Fig. \ref{pic1}). We write the electric field as ${\bf E} = -\nabla \phi$ so that the electric potential $\phi$ satisfies $\nabla^2 \phi = 0$ everywhere except for $z=0$ and $z=u$. Neglecting the effects of magnetism, we obtain the electrostatic Maxwell stress tensor, in component form, as:
$$\sigma_{ij} = \epsilon_0 \left (E_i E_j -{1\over 2} {\bf E}^2 \delta_{ij}\right ),$$
where $\epsilon_0 \approx 8.85\times 10^{-12} \ {\rm F/m}$ is the electric constant 
and $\delta_{ij}$ is the Kronecker delta. Here we are interested in the $zz$-component:
$$\sigma_{zz} = {\epsilon_0 \over 2} \left (E_z^2-{\bf E}_\parallel^2\right ), \eqno(1)$$
where ${\bf E}_\parallel = (E_x,E_y,0)$ is the parallel electric field.
We write the electric potential as:
$$\phi ({\bf x},t) = \int d^2q \ \phi ({\bf q},t) e^{i{\bf q}\cdot {\bf x}},$$
$$\phi ({\bf q},t) = {1\over (2 \pi)^2} \int d^2x \ \phi ({\bf x},t) e^{-i{\bf q}\cdot {\bf x}}.$$

Similarly we write:
$$\phi ({\bf x},t) = \int d^2q d\omega \ \phi ({\bf q},\omega) e^{i({\bf q}\cdot {\bf x}-\omega t)}.$$
To simplify the notation, in some of the equations below we will not write out the time (or $\omega$) dependency explicitly.
In the space between the surfaces the electric potential:
$$\phi = \int d^2q \left [\phi_1({\bf q}) e^{-qz}+ \phi_2({\bf q}) e^{qz} \right ] e^{i{\bf q}\cdot {\bf x}},$$
where ${\bf q} = (q_x,q_y)$ and ${\bf x} = (x,y)$ are 2D vectors. Thus for $z=0$:
$$E_z = \int d^2 q  \ q \left [ \phi_1({\bf q}) - \phi_2({\bf q}) \right ] e^{i{\bf q}\cdot {\bf x}}, \eqno(2)$$
and
$${\bf E}_\parallel = \int d^2 q  (-i{\bf q}) \left [\phi_1({\bf q}) + \phi_2({\bf q}) \right ] e^{i{\bf q}\cdot {\bf x}}.\eqno(3)$$
Using (1), (2) and (3), we have:
$$\int d^2 x \ \sigma_{zz} = 8 \pi^2 \epsilon_0 {\rm Re} \int d^2 q \ q^2 \phi_1({\bf q}) \phi^*_2({\bf q}).\eqno(4)$$

We now calculate $\phi_1({\bf q})$ and $\phi_2({\bf q})$. The electric potential for $z=u+d$ is denoted with $V({\bf x},t)$ and we define
$$V({\bf q},t) = {1\over (2 \pi )^2} \int d^2x \ V({\bf x},t) e^{-i{\bf q}\cdot {\bf x}},$$ 
$$V({\bf q},\omega) = {1\over (2 \pi )^3} \int d^2x dt \ V({\bf x},t) e^{-i({\bf q}\cdot {\bf x}-\omega t)}.$$ 

We write the electric potential $\phi({\bf q},z,\omega)$ as:
$$\phi = \phi_0 e^{qz} \ \ \ \ \ {\rm for}  \ \ z<0,$$
$$\phi = \phi_1 e^{-qz}+ \phi_2 e^{qz} \ \ \ \ \ {\rm for}  \ \ 0<z<u,$$
$$\phi = \phi_3 e^{-q(z-u-d)} + \phi_4 e^{q(z-u-d)} \ \ \ \ \ {\rm for}  \ \ u < z < u+d$$

Since $\phi$ must be continuous for $z=0$, $z=u$ and $z=u+d$ we get:
$$\phi_1+ \phi_2 = \phi_0, \eqno(5)$$
$$\phi_1 e^{-qu}+\phi_2 e^{qu} = \phi_3 e^{qd} +\phi_4 e^{-qd}, \eqno(6)$$
$$\phi_3+\phi_4 = V({\bf q},\omega). \eqno(7)$$

Let $\epsilon_1$ and $\epsilon_2$ be the dielectric permittivity of the lower and upper solid in Fig. \ref{pic1}, respectively. In our application the space between the bodies is filled with air with the dielectric permittivity of $\epsilon \approx 1$. From the boundary conditions $\epsilon_1 E_z(-0^+)= E_z(0^+)$ and  
$E_z(u-0^+)= \epsilon_2 E_z(u+0^+)$ we get:
$$-\phi_1+\phi_2 = \epsilon_1 \phi_0, \eqno(8)$$
$$-\phi_1 e^{-qu}+\phi_2 e^{qu} = \epsilon_2 \left (-\phi_3 e^{qd}+\phi_4 e^{-qd} \right ) . \eqno(9)$$

Using (5)-(9) we get
$$\phi_1 ({\bf q}, \omega) = 2 {1-\epsilon_1 \over 1+\epsilon_1} \cdot { V({\bf q},\omega)\over S(q,\omega)} , \eqno(10)$$
$$\phi_2 ({\bf q}, \omega) = 2 { V({\bf q},\omega)\over S(q,\omega)} , \eqno(11)$$
where
$$S (q,\omega)  = \left (e^{qd}+e^{-qd}\right ) \left ( e^{qu}+{1-\epsilon_1 \over 1+\epsilon_1} e^{-qu}\right )+$$
$$+ {1\over \epsilon_2} \left (e^{qd}-e^{-qd}\right ) \left (e^{qu} -{1-\epsilon_1 \over 1+\epsilon_1} e^{-qu}\right ) , \eqno(12)$$
where $\epsilon_1(\omega)$ and $\epsilon_2(\omega)$ in general are functions of $\omega$.

To summarize, in the most general case, from (4), (10) and (11) we have the electroadhesive force:
$$F(t) = 8 \pi^2 \epsilon_0 {\rm Re} \int d^2 q \ q^2 \phi_1({\bf q},t) \phi^*_2({\bf q},t), \eqno(13)$$
where
$$\phi_1 ({\bf q},t) = 2 \int_{-\infty}^\infty d \omega \ {1-\epsilon_1 (\omega) \over 1+\epsilon_1 (\omega)} \cdot 
{ V({\bf q},\omega)\over S(q,\omega)} e^{-i\omega t}, \eqno(14)$$
$$\phi_2 ({\bf q},t) = 2  \int_{-\infty}^\infty d \omega \ { V({\bf q},\omega)\over S(q,\omega)} e^{-i\omega t} . \eqno(15)$$

\vskip 0.2cm
{\bf 2.2 Mean-field theory of electroadhesion}

Real surfaces always have surface roughness, and the interfacial separation $u=u({\bf x},t)$ will vary with the spatial 
location ${\bf x}=(x,y)$ (see Fig. \ref{pic2}). When the adhesion pad is squeezed against a rough counter surface,
the pad will deform elastically, and the are of real contact, which determines or influences the friction force, will increase. 
Here we assume that the pad can be treated as a homogeneous material, and we neglect the influence of the conductive
strips on the elastic deformation of the pad material. This is a good approximation with respect to the surface roughness component with wavelengths shorter than the separation $d$ between the conductive strips and the pad surface. It is a good approximation for all surface roughness components if the conductive strips are very thin or if the strips have similar elastic properties as the pad material. The latter is the case if the electroadhesion pad is made from cPDMS as the conductive strips, and PDMS as the pad material\cite{[28]}.

In the so-called small-slope approximation, and assuming the surface roughness varies rapidly with the lateral coordinate compared to the electric potential $V({\bf x},t)$, the approach of Ref. \cite{[13],[15]} can be used to average over $u$, i.e., to perform an ensemble average over different realizations of the randomly rough surface. Note that the probability distribution of interfacial separations, $P(p,u)$, depends on $F(t)$ (the attractive force $F(t)$ pulls the surfaces into closer contact and hence reduce the surface separation $u({\bf x},t)$), but this can be taken into account using the mean-field approach described in Ref. \cite{[13],[15]}, and which we now briefly summarize. 

In the most general case we can write $V({\bf x},t) = V_0 f({\bf x},t)$, where $V_0$ is the amplitude of the applied voltage and $f({\bf x},t)$ a function which characterize how the electric potential $V({\bf x},t)$ varies in space and time. Using this and (13)-(15),
and averaging the electroadhesive  force over the distribution of interfacial separations, 
we can write force per unit surface area as:
$$p_{\rm a} = \langle \sigma_{zz} \rangle = V_0^2 \int_0^\infty P(p,u) G(p,u)\eqno (16)$$
where $p$ is the (nominal) pressure squeezing the solids together, and $P(p,u)$ the probability distribution of interfacial separation, 
and where $G(p,u)$ is the force $F(t)$ given by (13), divided by the nominal contact area $A_0$ and with $V_0^2$.

In the simplest approach one include the electrostatic attraction as a contribution to the external load. Thus we write the effective loading pressure as
$$p=p_0 + p_{\rm a}\eqno(17)$$
where $p_0$ is the (external) applied pressure. Intuitively, one expect this approach to be accurate 
when the interaction force between the surfaces is long-range, and a similar approach has been used 
for the attraction resulting from capillary bridges \cite{[21],[22]} and also in an earlier study of electroadhesion \cite{[13],[14],[15]}. Using (16) and (17) we have:
$$p= p_0 + V_0^2 \int_0^\infty du \ P(p,u) G(p,u),\eqno(18)$$

We can also write (18) as:
$$V_0^2 = {p-p_0 \over \int_0^\infty du \ P(p,u) G(p,u)} \eqno(19)$$
from which we can easily calculate $V_0$ as a function of the nominal contact pressure $p$.
Thus given $V_0$ and the applied (external) contact pressure $p_0$ the theory predict the electroadhesion
pressure $p$. Note that when $V_0=0$ then $p=p_0$ is equal to the external applied pressure $p_0$.
In the applications in Sec. 4 we have $p_0=0$.

To complete the theory we need the probability distribution $P(p,u)$. For randomly rough surfaces \cite{[23],[24],[25]}, for $u>0$ we have:
$$ P\left(p, u\right) \approx \frac{1}{A_{0}}\int_1^\infty d\zeta \ [-A^{\prime }(\zeta
)]\frac{1}{\left( 2\pi h_{\mathrm{rms}}^{2}(\zeta )\right) ^{1/2}}$$
$$\times \left[ \mathrm{exp}\left( -\frac{(u-u_{1}(\zeta ))^{2}}{2h_{\mathrm{%
rms}}^{2}(\zeta )}\right) +\mathrm{exp}\left( -\frac{(u+u_{1}(\zeta ))^{2}}{%
2h_{\mathrm{rms}}^{2}(\zeta )}\right) \right] .\eqno(20)$$
where $A(\zeta )$ is the (projected) contact area as a function of the magnification $\zeta$, and
$u_1(\zeta)$ is the separation between the surfaces in the area which moves out of contact when the magnification
increases from $\zeta$ to $\zeta+d\zeta$ (both quantities depend on the nominal contact pressure $p$). 
The quantity $h_{\rm rms} (\zeta)$ is the root-mean-square roughness
including only roughness components with the wavenumber $q> \zeta q_0$ i.e.
$$h_{\rm rms}^2 (\zeta)= \int_{q>\zeta q_0} d^2q \ C({\bf q}), \eqno(21)$$
where $C({\bf q})$ is the surface roughness power spectrum (see Ref. \cite{[23],[24],[25]} for more details).

In robotic applications the friction force between the adhesion pads and the counter surface is perhaps more important than the adhesion force. However, from the adhesion pressure $p$ one can calculate the adhesion force $F=pA_0$ and, if Coulomb friction law is valid, the friction force $F_{\rm f} = \mu F$, where $\mu$ is the friction coefficient. However, Coulombs friction law is only valid if the area of real contact $A$ is much smaller than the nominal contact area $A_0$. This is the case in most applications, but if the adhesion pad is made from an elastically very soft material like PDMS, and if the surfaces involved are very smooth, then the real contact area
may be similar to the nominal contact area. In this case, assuming elastic materials (no viscoelasticity), 
the friction force $F_{\rm f} = \tau_{\rm f} A$, where the frictional shear stress $\tau_{\rm f}$ 
must be determined experimentally, and where contact area $A$ is given (approximately) by the contact mechanics theory\cite{JCP}:
$${A \over A_0} = {\rm erf}\left ({1\over 2\surd G}\right ),\eqno(22)$$
where
$$G(\zeta) = {1 \over 8} \left ( {E\over (1-\nu^2)p}\right )^2 \xi \eqno(23)$$
where $$\xi^2 = 2 \pi \int_{q_{\rm L}}^{q_1} dq \ q^3 C(q)$$ is the surface mean-square slope. 
Using that ${\rm erf}(x) \approx 2x/\surd \pi$ for $x << 1$ one can show that when the squeezing force $F=p A_0$ is so small that $A<<A_0$, 
Eq. (22) reduces to $A/A_0 \approx \kappa p/(\xi E^*)$ where $E^*=E/(1-\nu^2)$ and $\kappa = (8/\pi )^{1/2} \approx 1.6$. 
Thus, for small (nominal) pressures $p$ the area of real contact $A$ is proportional to $p$; this is the physical basis for
Coulombs friction law for elastic solids.

\begin{figure}
\includegraphics[width=0.85\columnwidth]{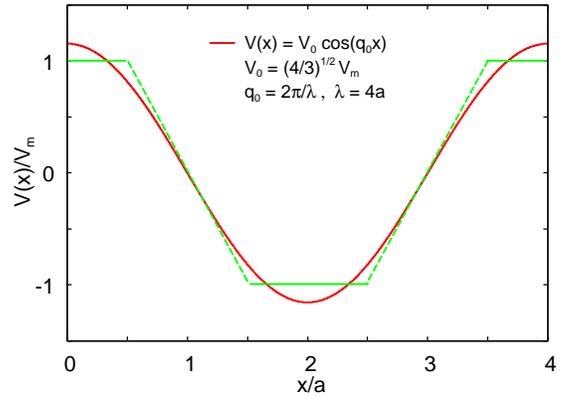}
\caption{\label{1x.2potential.eps}
The dependency of the electric potential on the spatial coordinate $x$.
In the experiment (see Ref. \cite{[27]}) an array of rectangular metallic strips was used, where the voltage
alternated between $V_{\rm m} = 1000 \ {\rm V}$ and $-1000 \ {\rm V}$. The metallic strips 
had the width $a=1 \ {\rm mm}$ and was separated by the distance $a$. The solid green line shows
the electric potential in the metallic strips and the dashed green line the approximate variation
in the potential between the strips. The red line is the potential used in the present modeling
with the amplitude $V_0 = (4/3)^{1/2} V_{\rm m}$ chosen so that the average of $V^2(x)$ is the same
in both cases.
}
\end{figure}

\vskip 0.3cm

{\bf 3 Case studies}

We consider three limiting cases of the theory. We first show that the $x$-dependency of the applied electric potential
in a realistic situation can be approximated with a cosines potential. Next we consider the case of perfectly flat surfaces
(Sec. 3.2) and the case of surface roughness but with a step-like turn on of the electric potential at time $t=0$ (Sec. 3.3).

\vskip 0.2cm
{\bf 3.1 Electroadhesion for a ${\rm cos}(q_0 x)$ electric potential}

Electroadhesion for robotic applications usually use a periodic array of metallic strips located close to the
adhesive pad surface. Thus in Ref. \cite{[27]}  Chen et al. performed electroadhesion 
experiments using an array ($300 \ {\rm mm} \times 240 \ {\rm mm}$) 
of rectangular copper electrodes. The metallic strips had the width $a=1 \ {\rm mm}$ and was separated by the distance $a$. 
The applied voltage alternated between $V_{\rm m} = 1000 \ {\rm V}$ and $-1000 \ {\rm V}$, as indicated by the green 
line in Fig. \ref{1x.2potential.eps}, which shows the dependency of the electric potential on the spatial coordinate $x$. 
In the figure the dashed green line is a linear interpolation which describes approximately the variation of the 
electric potential between the metallic strips. This potential can be described as a superposition 
of ${\rm cos} (qx)$ terms, but here we use the potential $V_0 {\rm cos} (q_0 x)$ given by the red 
line in Fig. \ref{1x.2potential.eps}. This potential has the same periodicity as the true potential, 
with the amplitude $V_0 = (4/3)^{1/2} V_{\rm m}$ chosen so that the average of $V^2(x)$ is the same in both cases.

Assume the external potential 
$$V({\bf x},t) = V(t) {\rm cos} (q_0 x) . \eqno(24)$$
In this case
$$V({\bf q},\omega) = V(\omega ) \ \delta (q_y) {1\over 2} \left [\delta (q_x+q_0)+\delta (q_x-q_0)\right ], \eqno(25)$$
and using that
$$V({\bf q},\omega) V^*({\bf q},\omega') = V (\omega )V^* (\omega' ) $$
$$\times [\delta (q_y)]^2 {1\over 4} \left ([\delta (q_x+q_0)]^2+[\delta (q_x-q_0)]^2\right )$$
$$= V (\omega )V^* (\omega' ) {A_0\over (4 \pi)^2} \delta (q_y) \left [\delta (q_x+q_0)+\delta (q_x-q_0)\right ],\eqno(26)$$
we get
$$F(t) =   \epsilon_0 A_0  q_0^2 \phi_1(q_0,t) \phi^*_2(q_0,t), \eqno(27)$$
where
$$\phi_1 (q_0,t) = 2 \int_{-\infty}^\infty d \omega \ {1-\epsilon_1 (\omega) \over 1+\epsilon_1 (\omega)} \cdot
{ V(\omega)\over S(q_0,\omega)} e^{-i\omega t} , \eqno(28)$$
$$\phi_2 (q_0,t) = 2  \int_{-\infty}^\infty d \omega \ { V(\omega)\over S(q_0,\omega)} e^{-i\omega t} . \eqno(29)$$

\vskip 0.2cm
{\bf 3.2 Electroadhesion for perfectly flat surfaces}

Assume now that the dielectric functions $\epsilon_1$ and $\epsilon_2$ are real and does not depend on the frequency $\omega$.
In this case $S(q,\omega) = S(q)$ is frequency independent and we get:
$$\phi_1 (q_0,t) = 2 {1-\epsilon_1  \over 1+\epsilon_1 } \cdot { V(t)\over S(q_0)} , $$
$$\phi_2 (q_0,t) = 2  {V(t)\over S(q_0)} , $$
and
$$F(t) =  q_0^2 4 \epsilon_0 A_0 {1-\epsilon_1  \over 1+\epsilon_1} \cdot { V^2(t)\over S^2(q_0)} . \eqno(30)$$

Assume that we have perfectly flat surfaces (no surface roughness). In this case $u=0$ and (12) reduces to
$$S={2\over 1+\epsilon_1} \left ( {\epsilon_2+\epsilon_1 \over \epsilon_2} e^{qd} +{\epsilon_2-\epsilon_1 
\over \epsilon_2} e^{-qd}\right ). \eqno(31)$$
If $\epsilon_1=\epsilon_2$
$$S(q,\omega)={1\over 1+\epsilon_1} e^{qd}\eqno(32) , $$
and
$$F(t) = - 4 \epsilon_0 A_0 (\epsilon_1^2 -1) q_0^2 e^{-2q_0 d} V^2(t) . \eqno(33)$$

This equation shows that strong electroadhesion require a large $q_0$, and that $2 q_0 d < 1$, or $d < \lambda_0/(4 \pi)$, where $\lambda_0 = 2 \pi /q_0$ is the wavelength of the applied voltage. For example, if $\lambda_0 = 10 \ {\rm \mu m}$ then the distance $d$ of the conducting strips to the surface of the adhesion pad should be at most $1  \ {\rm \mu m}$. In reality surface roughness will also exist, and since typically the amplitude of surface roughness is of micrometer order, it is clear that in practical applications $\lambda_0$ should be at least $10 \ {\rm \mu m}$ in order for the electric field to extend from the adhesive pad to the surface region of the counter material.

When optimizing an electro adhesive pad one must also take into account that electric breakdown can occur if the electric field strength
becomes too high. The critical electric field strength depends on the pad material, e.g., it is about $10^9 \ {\rm V/m}$
for PDMS\cite{[26],[27]}. When the distance $s$ between two metal strips (at different but fixed electric potential) decreases, the electric
field strength increases as $\approx 1/s$ so the breakdown voltage will decrease as we scale down the size of the array of conducting
strips.

\vskip 0.2cm
{\bf 3.3 Electroadhesion for a step voltage}

As a second example, assume that $V(t)=0$ for $t=0$ and $V(t)=V_0$ for $t>0$. In this case
$$V(\omega ) = {1\over 2 \pi i} \cdot {V_0 \over \omega+i0^+} , \eqno(34)$$
where $0^+$ is an infinite small positive number. 
We assume that the upper solid in Fig. \ref{pic1} is a perfect insulator so that $\epsilon_2$ a real frequency independent number. The lower solid is assumed to have a small electric conductivity. In this case
$$\epsilon_1 (\omega) = \epsilon_1^o + {i \gamma \over \omega} , \eqno(35)$$
where $\gamma = 1/(\epsilon_0 \rho)$ where $\rho$ is the electric resistivity. Using (12) we get:
$${2\over S} = {\alpha (1+\epsilon_1) \over \epsilon_1 +\beta} , \eqno(36)$$
$${1-\epsilon_1 \over 1+\epsilon_1}\cdot {2\over S} = {\alpha (1-\epsilon_1) \over \epsilon_1 +\beta} , \eqno(37)$$ 
where
$$\alpha = {2\over M-N} , \eqno(38)$$
$$\beta={M+N\over M-N} , \eqno(39)$$
where
$$M  = \left [\left (e^{qd}+e^{-qd}\right )+{1\over \epsilon_2}\left (e^{qd}-e^{-qd}\right )\right ]e^{qu} , \eqno(40)$$
$$N  = \left [\left (e^{qd}+e^{-qd}\right )-{1\over \epsilon_2}\left (e^{qd}-e^{-qd}\right )\right ]e^{-qu} . \eqno(41)$$

The integrals in (28) and (29) are now trivial to perform. Thus we have:
$$\phi_2 = {\alpha V_0 \over 2 \pi i} \int_{-\infty}^\infty d\omega \ {1 \over \omega+i0^+} \cdot {(1+\epsilon_1^o)\omega+i\gamma 
\over (\beta+\epsilon_1^o)\omega+i\gamma} e^{-i\omega t} . \eqno(42)$$
Closing the integration contour in the lower half of the complex $\omega$-plane gives
$$\phi_2 = \alpha V_0 \left [\left (1-e^{-t/\tau}\right )+\mu e^{-t/\tau}\right ] , \eqno(43)$$
where $\tau = (\beta+\epsilon_1^o)/\gamma$ and $\mu = (\epsilon_1^o+1)/(\epsilon_1^o+\beta)$.
Similarly we get:
$$\phi_1 = -\alpha V_0 \left [\left (1-e^{-t/\tau}\right )+\mu' e^{-t/\tau}\right ] , \eqno(44)$$
where $\mu' = (\epsilon_1^o-1)/(\epsilon_1^o+\beta)$, and hence
$$F(t) = - \epsilon_0 A_0 q_0^2 (\alpha V_0)^2 \left [\left (1-e^{-t/\tau}\right )+\mu e^{-t/\tau}\right ] $$
$$\times \left [\left (1-e^{-t/\tau}\right )+\mu' e^{-t/\tau}\right ] . \eqno(45)$$

\begin{figure}
\includegraphics[width=0.85\columnwidth]{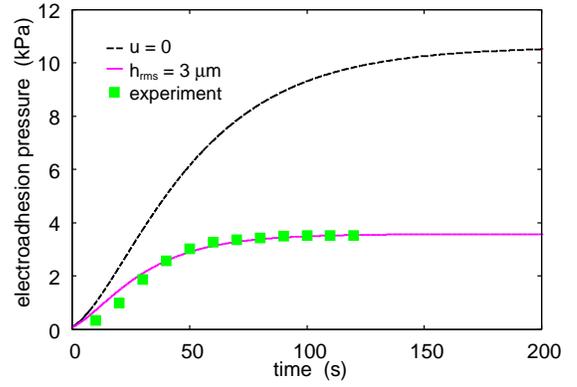}
\caption{\label{1time.2force1.eps}
The dependency of the electroadhesive pressure $p$ on time. The dashed and solid lines
are the theory prediction for perfectly smooth surfaces (i.e., complete contact, $u=0$),
and for a surface with the root-mean-square roughness $h_{\rm rms}=3 \ {\rm \mu m}$
(with the power spectrum shown in Fig. \ref{1logq.2logC.10.20.40mum.eps}, pink line), respectively.
The green squares are the measured
data for polyimide pad in contact with a silica glass surface (from \cite{[27]}). The glass surface has the
(measured) $\epsilon_1^o = 4.1$ and the electric resistivity $\rho = 10^{11} \ {\rm \Omega m}$.
The pad material is assumed to be a perfect insulator with $\epsilon_2 = 3$,
and the effective Young's modulus $E/(1-\nu^2) = 4.5 \ {\rm GPa}$. The adhesion pad
has a period distribution of rectangular conducting strips as described in Fig. \ref{1x.2potential.eps}.
}
\end{figure}

\begin{figure}
\includegraphics[width=0.85\columnwidth]{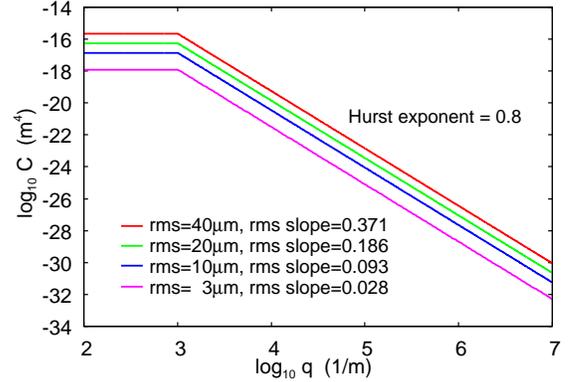}
\caption{\label{1logq.2logC.10.20.40mum.eps}
The surface roughness power spectra, used in the model calculations,
as a function of the wavenumber (log-log scale).
}
\end{figure}

\begin{figure}
\includegraphics[width=0.85\columnwidth]{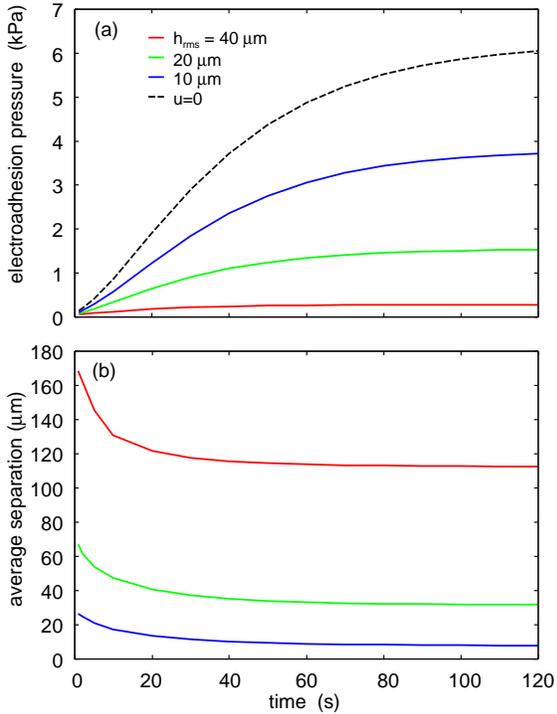}
\caption{\label{1time.2p.and.u.for.0mum.10mum.20mum.40mum.eps}
The time dependency of (a) the electroadhesive pressure $p$ and (b) the average surface separation.
The red, green and blue solid
lines are for surfaces with the root-mean-square roughness $40$, $20$ and $10 \ {\rm \mu m}$
with the power spectra shown in Fig. \ref{1logq.2logC.10.20.40mum.eps}.
The black dashed line is the theory prediction assuming the surface separation
with $u=0$. The substrate is assumed to
have $\epsilon_1^o = 4.1$ and the electric
resistivity $\rho = 10^{11} \ {\rm \Omega m}$ (as typical for silica glass).
The pad material is assumed to be a perfect insulator with $\epsilon_2 = 2.3$ and the Young's elastic modulus
$E=2 \ {\rm MPa}$ (as typical for PDMS). The adhesion pad
has a period distribution of rectangular
conducting strips as described in Fig. \ref{1x.2potential.eps}.
}
\end{figure}

\vskip 0.3cm
{\bf 4 Numerical results}

Let us compare the prediction of (45) with the experimental results presented in Ref. \cite{[17]}. 
Fig. \ref{1time.2force1.eps} shows the dependency of the electroadhesive pressure $p=F(t)/A_0$ on time. 
The dashed and solid lines are the theory prediction for perfectly smooth surfaces (i.e., complete contact, $u=0$),
and for a surface with the root-mean-square roughness $h_{\rm rms}=3 \ {\rm \mu m}$
(with the power spectrum shown in Fig. \ref{1logq.2logC.10.20.40mum.eps}, pink line), respectively.
The green squares are the measured 
data for the polyimide pad in contact with a silica glass surface (from \cite{[17]}). 
The glass surface has $\epsilon_1^o = 4.1$ and electric resitivity $\rho = 10^{11} \ {\rm \Omega m}$. 
The polyimide material is assumed to be a perfect insulator with $\epsilon_2 = 3$ and the effective Young's modulus $E/(1-\nu^2) = 4.5 \ {\rm GPa}$. 
The adhesion pad has a periodic distribution of rectangular conducting strips as described in Fig. \ref{1x.2potential.eps}.

No information about the surface roughness was given in Ref. \cite{[17]} but 
the rms roughness $h_{\rm rms}=3 \ {\rm \mu m}$ is typical for smooth
polymer surfaces. To apply the theory described in Sec. 2.2 one need the 
surface roughness power spectrum and the elastic modulus of the pad material, which was not 
given in Ref. \cite{[17]}. Here we have assumed that the surface is self-affine fractal like with
the Hurst exponent $H=0.8$, which is typical for real surfaces\cite{fractal}. The effective Young's elastic
modulus we use $E/(1-\nu^2) =4.5 \ {\rm GPa}$ is typical for polyimide.
We note that nearly the same result for the time dependency of the electroadhesive force, as found 
above using the full theory, result assuming the (fixed) surface separation $u=12 \ {\rm \mu m}$ in (45).

Polyimide has a rather high Young's elastic modulus ($E\approx 3.4 \ {\rm GPa}$) and is therefore not an ideal material
for adhesive pads unless both the pad surface and the substrate surface are very smooth and flat. 
A much better material for robotic applications is PDMS\cite{[28]}. 
Let us present some numerical results for this pad material.

Fig. \ref{1logq.2logC.10.20.40mum.eps}
shows the surface roughness power spectra, used in the model calculations, 
as a function of the wavenumber (log-log scale). (The power spectra in Fig. \ref{1logq.2logC.10.20.40mum.eps}  are typical for many surfaces,
see Ref. \cite{fractal}.)

In Fig \ref{1time.2p.and.u.for.0mum.10mum.20mum.40mum.eps}(a) and (b) we show the time-dependency of
the electroadhesive pressure $p$, and the average surface separation $u$. The blue, green and red solid
lines are for surfaces with the root-mean-square roughness $10$, $20$ and $40 \ {\rm \mu m}$,
with the power spectra shown in Fig. \ref{1logq.2logC.10.20.40mum.eps}.
The black dashed line in (a) is the theory prediction assuming the surface separation
with $u=0$. The substrate is assumed to 
have $\epsilon_1^o = 4.1$ and the electric 
resistivity $\rho = 10^{11} \ {\rm \Omega m}$ (as typical for silica glass).
The pad material is assumed to be a perfect insulator with $\epsilon_2 = 2.3$ and the Young's elastic modulus
$E=2 \ {\rm MPa}$ and Poisson ratio $\nu = 0.5$ (as typical for PDMS). Note that the electroadhesive pressure drops rapidly with increasing
surface roughness.

\vskip 0.3cm
{\bf 5 Summary and conclusion}

Soft adhesive pads based on electroadhesion are useful tools for shape-adaptive material handling of complex surfaces. 
In addition, they can be used to conform to roughness surfaces better than their rigid counterparts. 
In this paper, we have developed a general electroadhesion force model, assuming an elastic electroadhesion pad (made of two periodic arranged and coplanar 
conductive electrodes embedded in a soft dielectric) and a counter dielectric surface with a finite electrical conductivity. 
We have considered the most general case where the solids have arbitrary dielectric properties, the applied voltage has arbitrary time-dependency, and the contact surfaces have surface roughness.

We have considered in detail the limiting case of a ${\rm cos}(q_0 x)$ electric potential. 
Analytic and numerical results was presented for a step-voltage (in time). 
The numerical results was compared with the experimental data from Ref. \cite{[17]}, where a polyimide electroadhesion pad was in contact with a silica glass surface.
We have also presented a numerical results for a soft adhesive PDMS pad in contact with a glass substrate, with an applied (in time) step-voltage. 
We have varied the surface roughness and shown that the rougher the pad surface, the smaller the adhesive force and the quicker the adhesive 
pressure response. In addition, the electroadhesion force brings the PDMS surface closer to the 
substrate surface in a dynamic way. 

The model and numerical results presented in this work have the potential to 
fundamentally guide the optimization design of electroadhesion pads for various robotics tasks such as tool fixing, 
crawling/climbing, interconnecting, perching, anchoring, and material handling applications.

\end{document}